\documentclass{article}
\usepackage{hyperref}
\usepackage{graphicx}
\usepackage{subcaption}
\usepackage{amsmath}
\usepackage{amssymb} 
\usepackage{bm}  

\usepackage{tikz}        
\usetikzlibrary{shapes.geometric, arrows}
\usetikzlibrary{positioning}
\tikzstyle{rect} = [rectangle, draw, minimum width=2.5cm, minimum height=0.2cm]
\tikzstyle{elps} = [ellipse, draw, minimum width=2.5cm, minimum height=0.2cm]
\tikzstyle{arrow} = [thick,->,>=stealth]

\usepackage{booktabs}  
\usepackage{siunitx}   
\usepackage{algorithm}  
\usepackage{algpseudocode}
\usepackage{changepage} 
\usepackage{float}

\title{Sensitivity Analysis of State Space Models for \\ Scrap Composition Estimation in EAF and BOF}

\author{Yiqing Zhou, Karsten Naert and Dirk Nuyens}
\date{April 2025}

\begin{document}

\maketitle

\begin{abstract}
This study develops and analyzes linear and nonlinear state space models for estimating the elemental composition of scrap steel used in steelmaking, with applications to Electric Arc Furnace (EAF) and Basic Oxygen Furnace (BOF) processes. The models incorporate mass balance equations and are fitted using a modified Kalman filter for linear cases and the Unscented Kalman Filter (UKF) for nonlinear cases. Using Cu and Cr as representative elements, we assess the sensitivity of model predictions to measurement noise in key process variables, including steel mass, steel composition, scrap input mass, slag mass, and iron oxide fraction in slag. Results show that the models are robust to moderate noise levels in most variables, particularly when errors are below $10\%$. However, accuracy significantly deteriorates with noise in slag mass estimation. These findings highlight the practical feasibility and limitations of applying state space models for real-time scrap composition estimation in industrial settings.
\end{abstract}

\section{Introduction}
Steel is a foundational material in modern society, with applications ranging from construction and transportation to household appliances. However, steel production is a significant source of greenhouse gas emissions, presenting a major environmental challenge. As global demand for steel continues to rise, reducing the industry's environmental footprint has become increasingly urgent. One promising approach is the recycling of scrap steel, which can substantially lower energy consumption, reduce waste, and decrease carbon emissions. Nonetheless, integrating scrap steel into the production process remains complex due to the variable and unpredictable composition of scrap material.

The Basic Oxygen Furnace (BOF) and Electric Arc Furnace (EAF) are two common methods for producing steel, which can make use of recycled scrap. Although, using recycled scrap helps to reduce environmental impact, the unpredictable composition of scrap steel makes this a hard problem. Changes in the scrap's elemental fraction can cause issues with product quality, make the process harder to control, and complicate predictions about the process \cite{de2019basic, hay2021review}. 
If the uncertainty of scrap composition is high, only a limited portion of scrap can be used in the charge, requiring a greater reliance on iron ore and consequently increasing CO\textsubscript{2} emissions. 
Therefore, accurate estimation of the elemental fraction of different scrap types is essential for reducing this uncertainty. By enhancing the accuracy of scrap composition estimation, we can safely increase its usage in steelmaking and further reduce the industry's environmental footprint.

New techniques for modeling and improving these processes are helping to overcome these challenges. Methods such as least squares estimation \cite{sandberg2007scrap, birat2002quality}, maximum likelihood estimation \cite{arzpeyma2021model}, and stochastic optimization \cite{gaustad2007modeling} have been used to predict the amount of different elements in scrap and the uncertainty in those predictions. However, these methods often struggle to deal with changes in scrap composition over time, limiting their use in real-time control.

This study builds upon a novel framework that addresses these limitations through the use of state-space models and advanced filtering techniques. Specifically, Kalman filters and unscented Kalman filters are employed to estimate the evolving elemental composition of scrap steel in real-time, which can be found in~\cite{zhou2025model}. Rather than focusing on the detailed mathematical formulation of the models and algorithms, which have been thoroughly discussed in~\cite{zhou2025model}, this work emphasizes the sensitivity of estimation performance to measurement noise across different sensor inputs.

The remainder of this paper is organized as follows: Section~\ref{s:model} introduces a linear and a non-linear state-space model for two different types of elements, and corresponding filtering algorithms to fit the models.  Section~\ref{s:numerical_sensitivity} analyzes the sensitivity of estimation accuracy to measurement noise in various settings. Section~\ref{s:conclusion} concludes the paper with a summary of findings and potential directions for future work.

\section{State space models and corresponding filtering algorithms }
\label{s:model}
This section reintroduces the state space model and associated filtering algorithms presented in~\cite{zhou2025model}, with the aim of clarifying the modeling framework and the underlying assumptions.

\subsection{State space models}
In the steelmaking process, certain chemical elements only go into the steel, while others partition between the steel and the slag. The model for these two kinds of elements are different.
Various process variables are measurable in practice, such as the input mass of each scrap type, the total mass of tapped steel, and  the mass fraction of an element X of interest in the steel. Some other variables, including the mass of slag and the iron oxide content in slag, are often estimated using empirical models.

For elements that remain only in the steel, the observation model can be directly written as a linear function of the state variables, which represent the elemental composition of the scrap types. This results in a linear model.
However, for elements that partition between steel and slag, it is often challenging to measure or estimate all the necessary process variables, such as the elemental fraction in slag or the partition coefficients. To address this, the model is extended by introducing additional state variables to estimate the necessary process variables, allowing the observation to be expressed in a form consistent with the mass balance principles. Thus the observation is no longer a linear function of the state variable, which makes it a non-linear model.

\paragraph{Case 1: Element X retained in steel only.}
For an element \( X \) that enters only the steel, the system can be modeled using the following linear state space formulation for heats \( t = 1, 2, \ldots, \)
\begin{align}    
    \vec \alpha_{t+1} &= (1-\gamma)\,\vec\alpha_t + \gamma \, \vec \eta_t, 
    \quad
    &\vec \eta_t \sim \mathcal{D}(\vec q, Q),
    \label{eq:state_cu}\\
    y_t &= \vec m_t \cdot\vec \alpha_t + \varepsilon_t,
    &\varepsilon \sim \mathcal{N}(0, H_t).
    \label{eq:obs_cu}
\end{align}
The variables in this model are defined as follows.
\begin{itemize}
    \item \( \vec \alpha_t \): The fraction of element \( X \) across scrap types at heat \( t \); this is the unobserved state variable.
    \item \( \gamma \): A small positive constant representing temporal correlation between adjacent heats.
    \item \( \vec \eta_t \):  A random variable that introduces temporal variability into \( \vec \alpha_t \) and reflects the long-term average fraction of element \( X \) in each scrap type. Each component of $\vec\eta_t$ is sampled from a Beta distribution, with its mean given by the corresponding entry in \( \vec q \) and its variance determined by the corresponding diagonal entry in the diagonal matrix \( Q \).

    \item \( \vec m_t \): Input mass of each scrap type at heat \( t \).
    \item \( y_t \): Observed total mass of element \( X \) from scrap at heat \( t \). It can be calculate by other measurements using $ m_{{\rm steel}, t} f_{{\rm steel, X},t} - m_{{\rm hm}, t} f_{{\rm hm, X},t}$.
    \item \( \varepsilon_t \): Observation noise, modeled as zero-mean Gaussian noise with covariance \( H_t \).
\end{itemize}
Equation~\eqref{eq:state_cu} governs the state dynamics, while Equation~\eqref{eq:obs_cu} models the observations based on mass balance assumptions.

\paragraph{Case 2: Element X partitioning between steel and slag.}
For elements that are distributed across both steel and slag, the observation model becomes nonlinear due to the partition behavior. The extended state space model is given by
\begin{align}
    \vec \alpha^+_{t+1} 
    &= \begin{bmatrix} \vec \alpha_{t+1} \\ \vec c_{t+1} \end{bmatrix} 
    = (1 - \gamma) \begin{bmatrix} \vec \alpha_t \\ \vec c_t \end{bmatrix}
    + \gamma \begin{bmatrix} \vec \eta_t \\ \vec \theta_t \end{bmatrix},
    \quad \vec \eta_t 
     \sim \mathcal{D}(\vec q, Q), \quad \vec \theta_t \sim \mathcal{D}(\vec q_c, Q_c),
    \label{eq:state_cr} 
    \\
    y_t 
    &= 
        \frac{
            \vec m_t \cdot \vec \alpha_{t} + m_{\text{hm}, t} \, f_{\text{hm}, t}
        }{
            1 + (c_{1,t} + c_{2,t}  f_{\text{FeOn, slag}, t}) \, m_{\text{slag}, t} / m_{\text{steel}, t} 
        } + \varepsilon_t, 
     \quad \varepsilon \sim \mathcal{N}(0, H_t).
    \label{eq:obs_cr}
\end{align}
The new variables are listed below.

\begin{itemize}
    \item \( \vec \alpha_t^+ \): The extended state vector, with \( \vec c_t \) representing the partition coefficients for element \( X \).
    \item \( \vec \theta_t \): The corresponding random variable for $\vec c_t$,  where each component follows a Beta distribution with mean and variance be the corresponding component of vector $\vec q_c$ and of diagonal matrix $Q_c$.

    \item \( y_t \): Observed mass of element \( X \) in the steel, computed via a nonlinear mass balance formula.
    \item Other terms: mass of hot metal $m_{{\rm hm},t}$, fraction of X in hot metal $f_{{\rm hm}, t}$, fraction of iron oxide in slag $f_{{\rm FeOn, slag}, t}$, the mass of slag $m_{{\rm slag}, t}$, and the mass of steel $m_{{\rm steel}, t}$.
\end{itemize}
The state update equation~\eqref{eq:state_cr} is similar to~\eqref{eq:state_cu} for the linear model, while the observation equation~\eqref{eq:obs_cr} is no longer a linear combination of state variable $\vec \alpha_t^+$ anymore, which makes it a non-linear model.

\subsection{Filtering algorithms}
This section introduces filtering algorithms suitable for the linear and nonlinear state space models described above. The Kalman filter is used for the linear case, while the Unscented Kalman Filter (UKF) is adopted for the nonlinear model.

\paragraph{Linear Model: Kalman Filter}
For the linear state space model~\eqref{eq:state_cu}--\eqref{eq:obs_cu}, the Kalman filter provides a standard approach to estimate the state from noisy observations. It combines model predictions with measurements to yield updated state estimates over time.
Although \( \vec \eta_t \) follows a distribution \( \mathcal{D}(\vec q, Q) \) constrained within \([0,1]\), its small variance justifies approximating it as Gaussian. This leads to a slight modification of the standard Kalman filter, which typically assumes zero-mean noise~\cite[Section~4.3]{kalman1960approach}. In our case, the noise has nonzero mean \( \vec q \), altering the prediction step accordingly.
The parameters and initial state can be estimated using the Expectation-Maximization (EM) algorithm~\cite{moon1996expectation}, or some other numerical methods. However, since there are other measurements and professional knowledge available for the steelmaking process, it is recommended to manually tune the parameters.

\paragraph{Nonlinear Model: Unscented Kalman Filter}
For the nonlinear model~\eqref{eq:state_cr}--\eqref{eq:obs_cr}, we apply the Unscented Kalman Filter (UKF)~\cite{julier1997new, julier2004unscented}, following~\cite[Section~10.3]{durbin2012time}. The UKF uses the unscented transform to propagate a set of sigma points through nonlinear functions, capturing the posterior distribution more accurately than linear approximations.
Our model differs slightly from classical formulations~\cite[equations~(9.36)--(9.37)]{kalman1960approach} by allowing nonzero mean noise \( [\vec q, \vec q_c]^\top \), which requires small adaptations to the UKF algorithm. Hyperparameters are also tuned manually based on process knowledge.

\subsection{Input uncertainty analysis}
The model and associated filtering algorithms have been introduced and motivated in~\cite{zhou2025model}, which also includes a detailed analysis of model misspecification and supporting numerical results. However, \cite{zhou2025model} assumes that all input variables are measured precisely, accounting only for observation noise in the form of \( \varepsilon_t \) in~\eqref{eq:obs_cu} and~\eqref{eq:obs_cr}.

In the following section, we extend this analysis by examining the impact of measurement noise in other variables, such as the mass of input scrap and the mass of steel, on the accuracy and robustness of state estimation and model prediction.

\section{Numerical results}
\label{s:numerical_sensitivity}

Validation of the algorithms, the impact of model misspecification, and results based on real data are discussed in~\cite{zhou2025model}. This section focuses on evaluating the sensitivity of model predictions with respect to various measurements. The linear model~\eqref{eq:state_cu}--\eqref{eq:obs_cu} is demonstrated using Cu as a representative element, while Cr is used for the nonlinear model~\eqref{eq:state_cr}--\eqref{eq:obs_cr}.\

The models rely on several key measurements:
\begin{itemize}
    \item steel mass \( m_{{\rm steel}, t} \),
    \item element fraction in steel \( f_{{\rm steel, X}, t} \),
    \item input mass of scrap types \( \vec m_t \),
    \item slag mass \( m_{{\rm slag}, t} \),
    \item iron oxide fraction in slag \( f_{{\rm FeOn, slag}, t} \).
\end{itemize}

To evaluate the robustness of the models under realistic measurement uncertainty, we introduce synthetic noise into key process variables using a multiplicative Gaussian noise model
\[
    \tilde{m} = (1 + \xi) \, m, \quad \xi \sim \mathcal{N}(0, c),
\]
with noise levels \( c \in [0\%,\, 20\%] \). Since noise in \( m_{{\rm steel}, t} \) and \( f_{{\rm steel, X}, t} \) affects only the observation model~\eqref{eq:obs_cu}, their effect can be absorbed into the observation noise covariance \( H_t \) and is not analyzed further.

\subsection{Synthetic data generation}
The model was tested on actual BOF data from ArcelorMittal in the context of~\cite{zhou2025model}. Our model has also been run on real EAF data from ArcelorMittal in the context of this paper; however, due to confidentiality constraints, those results cannot be published. Therefore, the results presented in this paper are based on synthetic, yet realistic, data.

To enable a controlled and interpretable analysis, we now generate synthetic data that closely mimics realistic operating conditions. This also allows for a more systematic sensitivity analysis, as the true values of all variables are known.
To conduct the sensitivity analysis, synthetic data is generated under controlled conditions. We assume the presence of $45$ distinct scrap types and $T=20000$ heats. The following experiments are designed to mimic BOF process data. According to the models~\eqref{eq:state_cu}--\eqref{eq:obs_cu} and~\eqref{eq:state_cr}--\eqref{eq:obs_cr}, the EAF process can be represented as a special case by setting \( m_{\rm hm} = 0 \).

\paragraph{Elemental fraction of scrap types}
For both Cu and Cr, we first generate the initial elemental fraction vector \( \vec \alpha_1 = \vec q \). The $45$ scrap types are divided into three groups (15 types per group), with their values sampled from the following uniform distributions with unit [ppm].
\begin{itemize}
    \item Group 1 (types 1–15): \( \mathcal{U}(200, 1000) \).
    \item Group 2 (types 16–30): \( \mathcal{U}(1000, 2000) \).
    \item Group 3 (types 31–45): \( \mathcal{U}(2000, 5000) \).
\end{itemize}
This grouping reflects common practices in real production,  where scrap types vary in impurity levels, some containing more Cu and Cr than others. 
The covariance matrix \( Q \) is diagonal, with elements \( Q_{i,i} = 5 q_i^2 \) for \( i = 1, \ldots, 45 \). The temporal correlation constant is set to \( \gamma = \ln(2) / 1000 \). The sequence \( \vec \alpha_t \) for \( t > 1 \) is then generated recursively using the linear state update~\eqref{eq:state_cu} or the nonlinear model~\eqref{eq:state_cr}, depending on the element. These choices are motivated in~\cite{zhou2025model}.

\paragraph{Partition coefficients of Cr}
For the Cr model, the initial and prior mean of the partition coefficients is set to \( \vec c_1 = \vec q_c = [9.7,\ 0.01]^\top \). The covariance matrix \( Q_c \) is diagonal: $Q_c$ is set to $[5\times9.7^2, \, 0; \, 0,\, 5 \times0.01^2]$.
The values of \( \vec c_t \) for \( t > 1 \) are then generated recursively via the state equation~\eqref{eq:state_cr}.

\paragraph{Input mass of scrap types}
We generate a matrix for input mass of scrap types, where the $t$-th row of the matrix represents the value of $\vec m_t$, $t = 1,\ldots, T$. The matrix satisfies some conditions: all the elements are non-negative, and the summation of each row is around $70\,$t. This row sum reflects the practical capacity constraints of a BOF, making it a plausible and realistic.
The details about generating the matrix will be discussed in Section~\ref{sss:impact_fix_recipe}.

\paragraph{Other variables}
Table~\ref{tab:bound_uniform} lists the variables whose values are sampled from uniform distributions, along with their respective upper and lower bounds.

\begin{table}[htbp]
\centering
\caption{The lower and upper bound of the uniform distribution for variables.}
\label{tab:bound_uniform}
\begin{tabular}{SSSS}
    \toprule
    {Variables} & {Lower} & {Upper} & {Unit}\\
    \midrule
    {$m_{{\rm hm}, t}$} & {$280$}   & {$ 290$} & {[t]}\\
    {$f_{{\rm hm, Cu}, t}$} & {$20$}   & {$ 30$} & {[ppm]}\\
    {$f_{{\rm hm, Cr}, t}$} & {$50$}   & {$ 200$} & {[ppm]}\\
    {$m_{{\rm slag}, t}$} & {$20$}   & {$ 30$} & {[ppm]}\\
    {$m_{{\rm steel}, t}$} & {$340$}   & {$ 350$} & {[ppm]}\\
    {$f_{{\rm FeOn, slag}, t}$} & {$20$}   & {$ 30$} & {[ppm]}\\
    \bottomrule
\end{tabular}
\end{table}

\paragraph{Elemental fraction of steel}
Since all of the other variables have already been generated, the elemental fraction of steel can be calculated using mass balance. And the standard deviation of measurement error for $f_{{\rm steel, Cu}, t}$  and $f_{{\rm steel, Cr}, t}$ are $10$ and $5\,$ppm, respectively. 

In the following experiments, we evaluate both the prediction error for the Cu and Cr fractions in steel and the estimated elemental fractions of the scrap types. This means that, in an ideal scenario, the prediction error would have zero mean and a standard deviation of $10\,$ppm for Cu and $5\,$ppm for Cr, representing the theoretical best-case performance under the assumed model.

\subsection{Input scrap mass}
\subsubsection{Impact of fixed scrap recipes}
\label{sss:impact_fix_recipe}
In industrial steelmaking, it is common for plants to reuse predefined ``recipes'' of scrap combinations tailored for specific steel grades. These recipes often consist of fixed proportions of several scrap types, resulting in limited variability in the input mass vectors \( \vec m_t \) over time. Mathematically, this lack of diversity is reflected in the matrix \( M \), constructed by stacking all \( \vec m_t \) vectors across timesteps,  which often becomes rank-deficient or exhibits a high condition number.

To investigate the implications of this structural limitation, we conduct numerical experiments by generating matrices with different properties, while keeping the total scrap input mass similar to the typical value of $70\,$t. There are different ways of constructing the matrix of $\vec{m}_t$ values corresponding to different scenarios that will be used in the experiments. The following ways will be used.
\begin{itemize}
    \item \textbf{Full-rank matrix with controlled condition number:} This matrix is generated using singular value decomposition to achieve a specific condition number. In practice, fixed scrap recipes can result in such a matrix structure.
    \item \textbf{Low-rank matrix:} A low-rank structure is imposed by multiplying two random matrices of sizes $m \times r$ and $r \times n$.  Since rank-deficient matrices have an infinite condition number, they represent a case when fixed recipes are used with very little variability.
    \item \textbf{Sparse matrix:} This matrix is constructed to be sparse, with each row containing at least one non-zero element and the overall density controlled by a specified parameter. After initial sparsification, additional mass is selectively added to a few specific scrap types (e.g., columns 1, 23, and 45), simulating the practical preference for well-characterized and frequently used scrap types. 
\end{itemize}    
For all of these methods each row is finally normalized to have a total mass of approximately $70\,$t. This setup is quite common in real applications, as scrap types with lower uncertainty are often favored in production planning.

\begin{figure}[htbp]
    \centering
    \begin{subfigure}[b]{0.45\linewidth}
        \includegraphics[width=\linewidth]{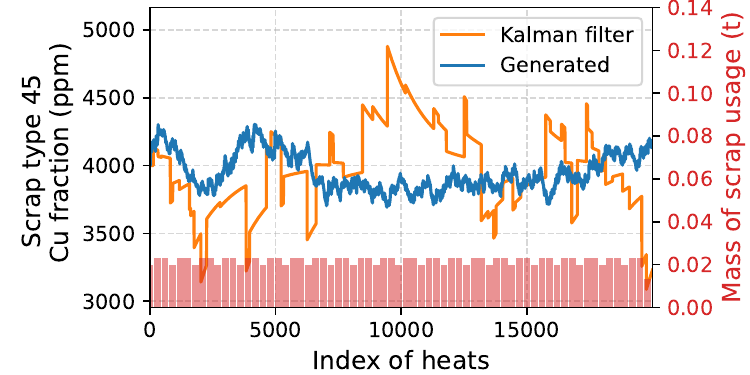}
        \caption{\centering Identity-based matrix rank = $45$, condition number = $1$.}
    \end{subfigure} 
    \begin{subfigure}[b]{0.45\linewidth}
        \includegraphics[width=\linewidth]{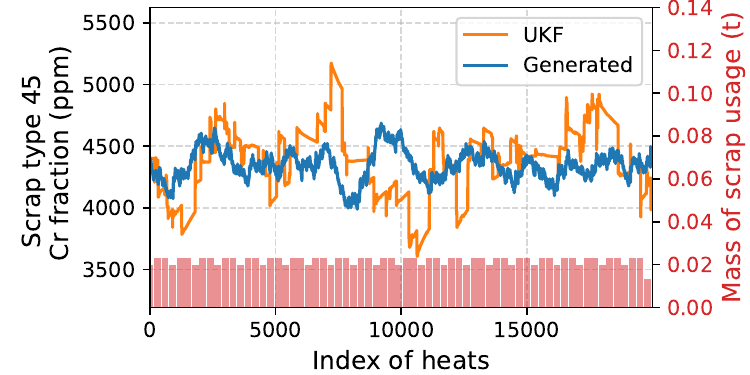}
        \caption{\centering Identity-based matrix rank = $45$, condition number = $1$.}
    \end{subfigure} 
    \\    
    \begin{subfigure}[b]{0.45\linewidth}
        \includegraphics[width=\linewidth]{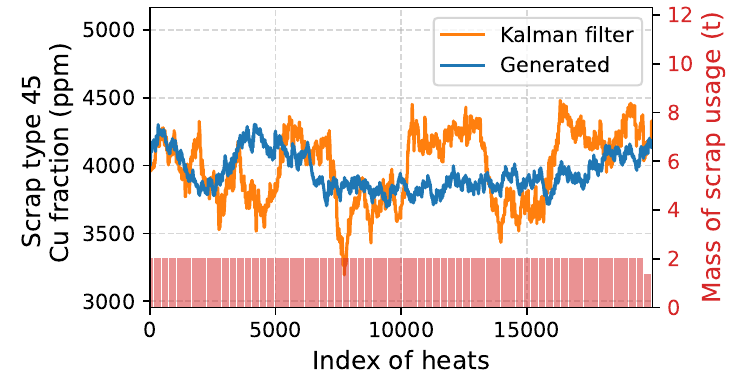}
        \caption{\centering Low rank matrix, rank = $20$.}
    \end{subfigure}
    \begin{subfigure}[b]{0.45\linewidth}
        \includegraphics[width=\linewidth]{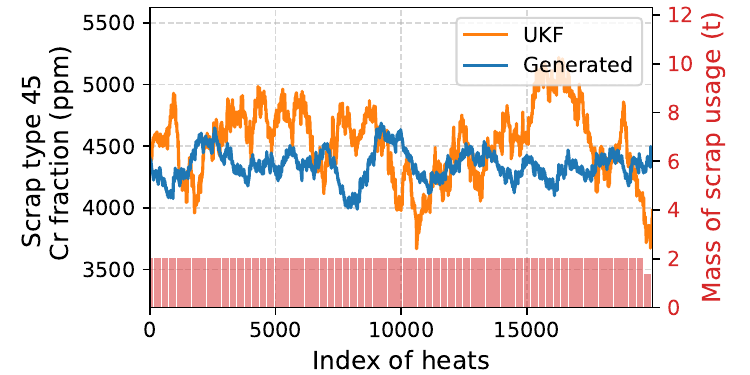}
        \caption{\centering Low rank matrix, rank = $20$.}
    \end{subfigure}
    \\    
    \begin{subfigure}[b]{0.45\linewidth}
        \includegraphics[width=\linewidth]{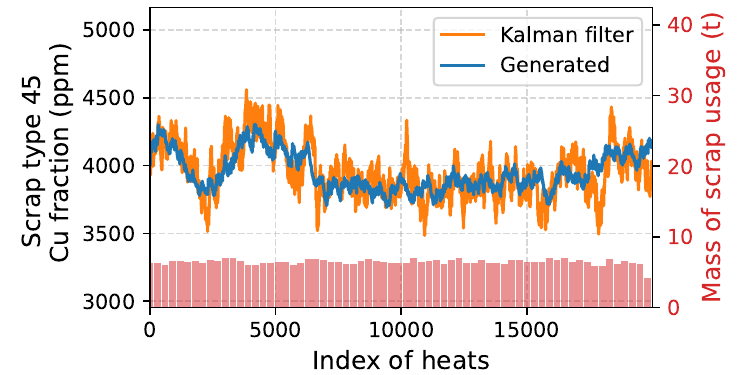}
        \caption{\centering Sparse matrix, rank = $45$.}
    \end{subfigure}
    \begin{subfigure}[b]{0.45\linewidth}
        \includegraphics[width=\linewidth]{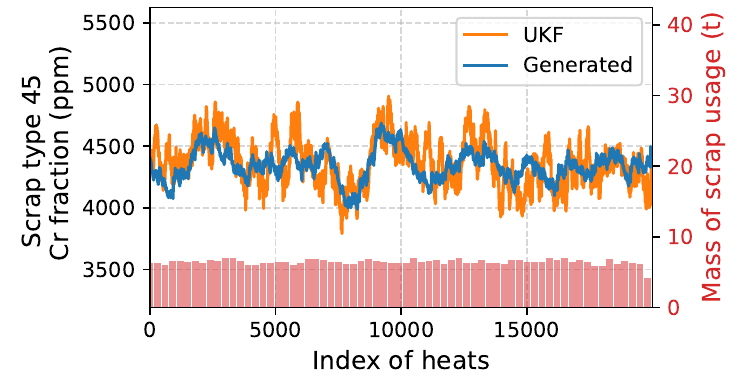}
        \caption{\centering Sparse matrix, rank = $45$.}
    \end{subfigure}
    \caption{ Effects of rank and condition number of matrix $M$ about input scrap mass on the estimated element fraction of scrap type $45$.
    The left column is the results about Cu and the right column shows the results about Cr. The last row performs the best in estimating the scrap composition, since it is used frequently and with more quantities.
    }  
    \label{fig:sens_m_scrap_rank}
\end{figure}

The results are summarized in Table~\ref{tab:sens_m_scrap_rank} and visualized in Figure~\ref{fig:sens_m_scrap_rank}.
Each row in the table corresponds to a different matrix $M$. The first row of this table corresponds to a case where only one scrap type is used per heat and, although unrealistic, serves as a best possible case to which the other results can be compared.
The second row uses a sparse matrix, and the third row uses a full rank matrix with high condition number. The fourth row uses a low rank matrix.
The table reports the mean and standard deviation of the prediction error for elemental fractions in steel. The standard deviation of observing $f_{{\rm steel, Cu}, t}$ and $f_{{\rm steel, Cr}, t}$ are $10\,$ppm and $5\,$ppm, respectively, representing an approximate lower bound on achievable error. The results in the table approach or even surpass this bound in some cases. 

However, a closer look at the figures reveals more nuanced behavior. The red bars show the average input mass of the scrap type grouped over $30$ heats. In the first two rows, the scrap type is used frequently but in small quantities. In contrast, in the last row, the scrap type is used consistently and in larger quantities. This provides more information for estimating its composition. Thus the estimated elemental fraction tracks the generated data much better.

These observations suggest an important practical insight: smaller condition number does not necessarily help to estimate the elemental fraction of scrap precisely, as shown in the first row in Figure~\ref{fig:sens_m_scrap_rank}.  When scrap types appear only in fixed ratios or low amounts, the model struggles to distinguish their individual contributions, relying more heavily on priors and causing estimates to revert toward the prior mean. Therefore, in practice, to improve the estimation of a particular scrap type’s elemental fraction, it is advisable to use that scrap type more frequently and in larger amounts.

\begin{table}[htbp]
\centering
\caption{Effect of matrix rank and condition number on predicted Cu and Cr fractions in steel (ppm). The standard deviation of measurement error for $f_{{\rm steel, Cu}, t}$ and  $f_{{\rm steel, Cr}, t}$ are $10\,$ppm and $5\,$ppm, respectively, representing the theoretical best-case performance.}
\label{tab:sens_m_scrap_rank}
\resizebox{\textwidth}{!}{
\begin{tabular}{cccccc}
    \toprule
    {${\rm Rank}\,(M)$} & {${\rm Cond}\,(M)$} & {${\rm E}[\varepsilon(f_{{\rm steel, Cu}, t})]$} & {${\rm \sigma}[\varepsilon(f_{{\rm steel, Cu}, t})]$} & {${\rm E}[\varepsilon(f_{{\rm steel, Cr}, t})]$} & {${\rm \sigma}[\varepsilon(f_{{\rm steel, Cr}, t})]$} \\
    \midrule
    {$45$ (Identity-based matrix)} & {$1$}   & {$ -0.016 $} & {$ 0.72 $}  & {$ -0.15 $}  &  {$ 2.23 $} \\ % done
    {$45$ (Sparse matrix)} & {$25$}  & {$ -0.33 $} & {$ 10.79$}  & {$ -0.07 $}  &  {$ 5.39 $} \\% done
    {$45$ (Full rank matrix)} & {$719104$}  & {$ 0.75 $} & {$ 7.59 $}  & {$ -0.12 $}  &  {$ 3.63 $} \\% done
    {$ 20 $ (Low rank matrix)} & {$\infty$}   & {$1.50 $} & {$ 7.73 $}  & {$ -0.28 $}  &  {$ 3.38 $} \\% done
    \bottomrule
\end{tabular}
}
\end{table}

Since the sparse matrix almost reaches the theoretical best-case performance in Table~\ref{tab:sens_m_scrap_rank} and also tracks the generated true elemental fraction well in Figure~\ref{fig:sens_m_scrap_rank}, the following experiments are all implemented with the same matrix.

\subsubsection{Sensitivity to measurement noise}
\begin{figure}[htbp]
    \centering
    \begin{subfigure}[b]{0.45\linewidth}
        \includegraphics[width=\linewidth]{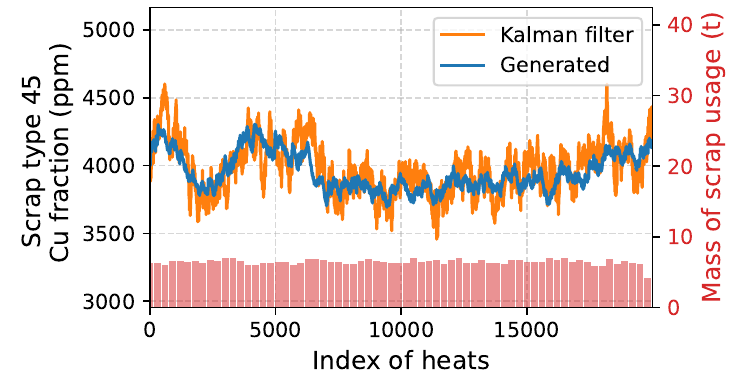}
        \caption{$1\%$ measurement noise on $\vec m_t$.}
    \end{subfigure} 
    \begin{subfigure}[b]{0.45\linewidth}
        \includegraphics[width=\linewidth]{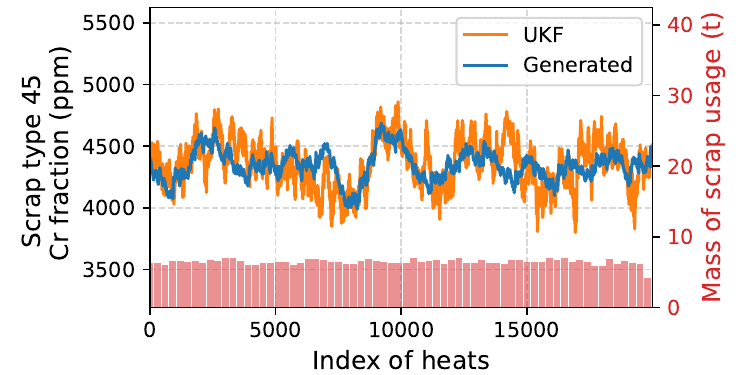}
        \caption{$1\%$ measurement noise on $\vec m_t$.}
    \end{subfigure} 
    \\
    \begin{subfigure}[b]{0.45\linewidth}
        \includegraphics[width=\linewidth]{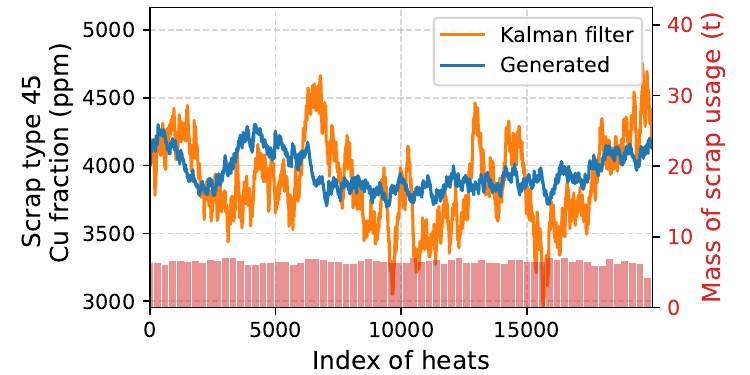}
        \caption{$10\%$ measurement noise on $\vec m_t$.}
    \end{subfigure} 
    \begin{subfigure}[b]{0.45\linewidth}
        \includegraphics[width=\linewidth]{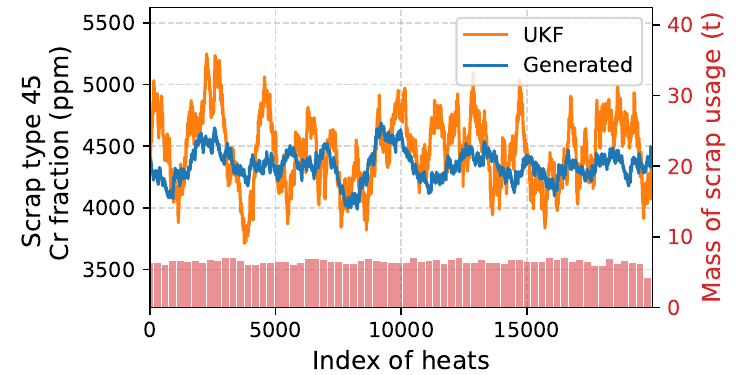}
        \caption{$10\%$ measurement noise on $\vec m_t$.}
    \end{subfigure} 
    \caption{Effects of measurement noise on input scrap mass on the estimated Cu and Cr fraction of scrap type $45$.
    The left column is the results about Cu and the right column shows the results about Cr. 
    The red bars in the figure tell the usage of the scrap type.
    }    
    \label{fig:KF_cucr80_f_soort}
\end{figure}

Accurate measurement of the input scrap mass \( \vec m_t \) is essential for estimating the elemental fraction of the steel and scrap. However, in practice, errors may occur due to handling losses, inaccurate weighing, or manual loading deviations. These uncertainties can propagate through the model and affect the quality of the composition estimates.
To evaluate the model's robustness, we introduce multiplicative Gaussian noise to \( \vec m_t \) and assess its effect on the predicted elemental fractions in the steel and scrap types. The results are summarized in Table~\ref{tab:sens_m_scrap} and visualized in Figure~\ref{fig:KF_cucr80_f_soort}.

The findings indicate that, when the measurement error of input scrap mass get larger, the bias and standard deviation of error of predicted elemental fraction in steel also gets larger. 
The model is relatively robust to small levels of noise. When the noise level remains below \( 20\% \), both the bias and standard deviation of the predicted Cu and Cr fractions in steel remain low and stable. 
For estimating the elemental fraction of scrap, the results are influenced more, compared with the last row of Figure~\ref{fig:sens_m_scrap_rank}. 
These results demonstrate that the model is robust to realistic levels of measurement noise. Even with noise levels up to \( 20\% \), the accuracy for predicting both Cu and Cr fraction in steel remains largely unaffected, with minimal increases in bias or variance, while the estimated Cu and Cr fraction of scrap types are affected. 

This suggests that the model can reliably handle the typical uncertainties encountered in industrial scrap handling and weighing, provided that measurement errors are reasonably controlled. Such robustness is particularly valuable for real-time monitoring and decision-making in steelmaking processes, where perfect data quality cannot always be guaranteed.

\begin{table}[htbp]
\centering
\caption{Effect of measurement noise in scrap input mass on predicted Cu and Cr fractions in steel (ppm). The first row is the same as the second row in Table~\ref{tab:sens_m_scrap_rank}, representing the theoretical best-case results. Since the true value of $f_{{\rm steel, Cu}, t}$ range between $200$ to $600\,$ppm, and $f_{{\rm steel, Cr}, t}$ ranges between $200$ to $400\,$ppm, all results reported in the table are considered acceptable.}

\label{tab:sens_m_scrap}
\resizebox{\textwidth}{!}{
\begin{tabular}{SSSSS}
    \toprule
    {${\rm noise}\,(\vec m_t)$} & {${\rm E}[\varepsilon(f_{{\rm steel, Cu}, t})]$} & {$\sigma[\varepsilon(f_{{\rm steel, Cu}, t})]$} & {${\rm E}[\varepsilon(f_{{\rm steel, Cr}, t})]$} & {$\sigma[\varepsilon(f_{{\rm steel, Cr}, t})]$} \\
    \midrule
    {$ 0\%  $}      & {$ -0.33 $} & {$ 10.79$}  & {$ -0.07 $}  &  {$ 5.39 $} \\
    {$ 1\%  $}      & {$ 0.23 $} & {$ 10.47 $}  & {$ -0.12 $}  &  {$ 5.33 $} \\
    {$ 5\%  $}      & {$ 0.32 $} & {$ 13.30 $}  & {$ 0.44 $}  &  {$ 7.13 $} \\
    {$ 10\% $}      & {$ -1.52 $} & {$ 16.69 $}  & {$ -0.18 $}  &  {$ 9.31 $} \\
    {$ 20\% $}      & {$ -3.21 $} & {$ 19.25 $}  & {$ -1.26 $}  &  {$ 12.88$} \\
    \bottomrule
\end{tabular}
}
\end{table}

\subsection{Slag mass}
Slag mass \( m_{{\rm slag}, t} \) plays a critical role in the nonlinear model, particularly for elements such as Cr that partition between steel and slag. In practical settings, however, slag mass is not directly measured but estimated from other process variables or inferred using empirical models, making it inherently uncertain. This makes it important to understand how sensitive the model is to potential measurement errors in this quantity.

\begin{figure}[htbp]
    \centering
    \begin{subfigure}[b]{0.45\linewidth}
        \includegraphics[width=\linewidth]{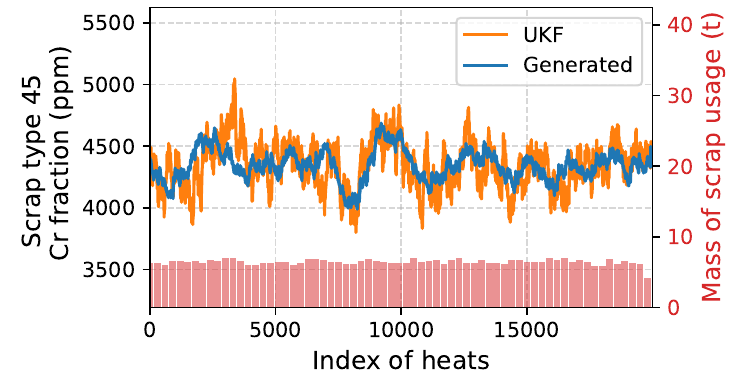}
        \caption{$1\%$ measurement noise on $ m_{{\rm slag}, t}$.}
    \end{subfigure} 
    \begin{subfigure}[b]{0.45\linewidth}
        \includegraphics[width=0.7\linewidth]{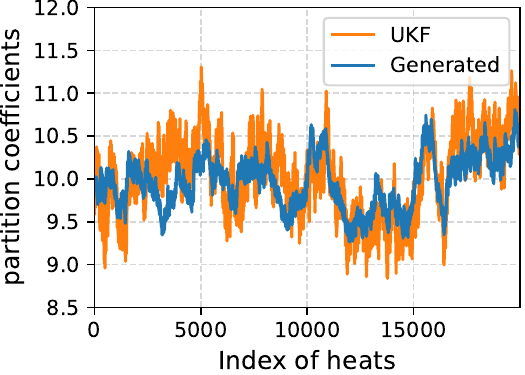}
        \caption{$1\%$ measurement noise on $ m_{{\rm slag}, t}$.}
    \end{subfigure} 
    \\
    \begin{subfigure}[b]{0.45\linewidth}
        \includegraphics[width=\linewidth]{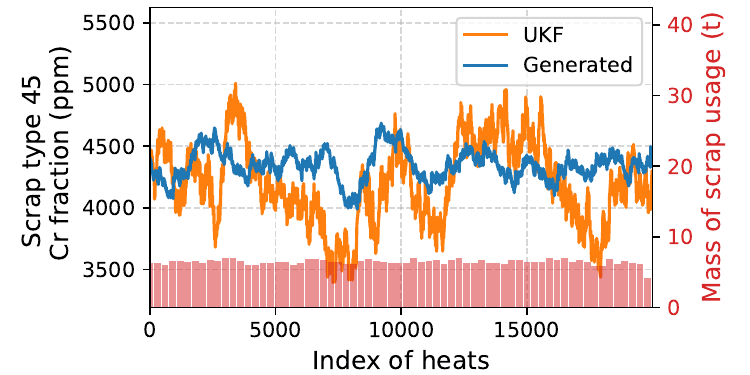}
        \caption{$10\%$ measurement noise on $ m_{{\rm slag}, t}$.}
    \end{subfigure}
    \begin{subfigure}[b]{0.45\linewidth}
        \includegraphics[width=0.7\linewidth]{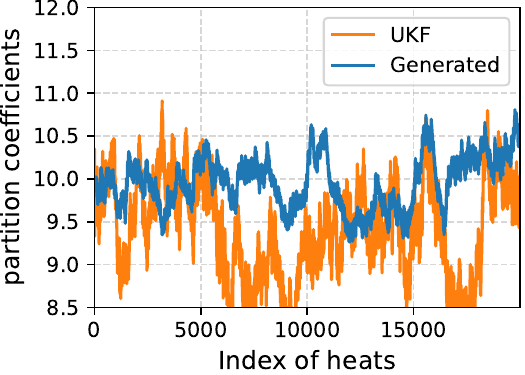}
        \caption{$10\%$ measurement noise on $ m_{{\rm slag}, t}$.}
    \end{subfigure}
    \caption{Effects of measurement noise on slag mass on the estimated Cr fraction of scrap type 80 and partition coefficients.
    The left columns are the results about Cr fraction of scrap type $45$ and the right column shows the results about partition coefficients. The upper row shows the results with $1\%$ measurements noise and the bottom row shows the results of $10\%$ measurement noise of input scrap mass.
    }    
    \label{fig:KF_cucr80_f_slak}
\end{figure}

Table~\ref{tab:sens_m_slag} and Figure~\ref{fig:KF_cucr80_f_slak} present the results of introducing controlled noise into the slag mass and evaluating its impact on the estimation of Cr content. Cu is not affected in this analysis, as slag-related variables are not part of the linear Cu model.
The results show that the model maintains reasonable robustness to noise up to \( 20\% \), with only slight changes in bias and variance. However, the error of estimated Cr fraction of scrap gets larger when the noise level reaches \( 10\%\).

These findings highlight that among all the key process variables, the slag mass is one of the most influential in the nonlinear model. Accurate estimation of slag mass is thus essential for reliable prediction of steel composition and estimation of scrap composition. In industrial practice, improving the accuracy of slag mass estimation, through more direct measurements, better sensors, or refined empirical models, could significantly enhance the reliability of data-driven state estimation models like the one presented in this study.

\begin{table}[htbp]
\centering
\caption{Effect of slag mass noise on predicted Cr fraction in steel (ppm).  The first row is the same as the second row in Table~\ref{tab:sens_m_scrap_rank}, representing the theoretical best-case results. Since the true value of $f_{{\rm steel, Cr}, t}$ ranges between $200$ to $400\,$ppm, all results reported in the table are considered acceptable.}
\label{tab:sens_m_slag}
\begin{tabular}{SSS}
    \toprule
    {${\rm noise}\,(\vec m_{\rm slag})$} & {${\rm E}[\varepsilon(f_{{\rm steel, Cr}, t})]$} & {$\sigma[\varepsilon(f_{{\rm steel, Cr}, t})]$} \\
    \midrule
    {$ 0\%  $}     & {$ -0.07 $}  &  {$ 5.39 $} \\
    {$ 1\%  $}     & {$ -0.01 $}  &  {$ 5.44 $} \\
    {$ 5\%  $}     & {$ 0.52 $}  &  {$ 7.21 $} \\
    {$ 10\% $}     & {$ 0.12 $}  &  {$ 9.21 $} \\ 
    {$ 20\% $}     & {$ -0.07 $}  &  {$ 13.06 $} \\
    \bottomrule
\end{tabular}
\end{table}

\subsection{Iron oxide fraction in slag}
The iron oxide fraction in slag, denoted as \( f_{{\rm FeOn, slag}, t} \), is a key input for estimating the partition coefficient \( \ell_X \), which governs the distribution of certain elements between steel and slag. In the model, this relationship is captured through a linear approximation of the form $\ell_t = c_{1,t} + c_{2,t} f_{{\rm FeOn, slag}, t}$, where the parameters \( c_{1,t} \) and \( c_{2,t} \) are fitted coefficients. This was already successfully tested on real BOF data in~\cite{zhou2025model}.

Because this partition coefficient is modeled as a function of \( f_{{\rm FeOn, slag}, t} \), any error in measuring this variable could potentially propagate into the estimation of element concentrations. However, the use of a flexible linear parameterization allows the model to partially absorb or correct for such errors through the adjustment of \( c_{1,t} \) and \( c_{2,t} \). As a result, the model is expected to exhibit some robustness to moderate inaccuracies in this measurement.

\begin{figure}[htbp]
    \hfill
    \begin{subfigure}[b]{0.45\linewidth}
        \includegraphics[width=\linewidth]{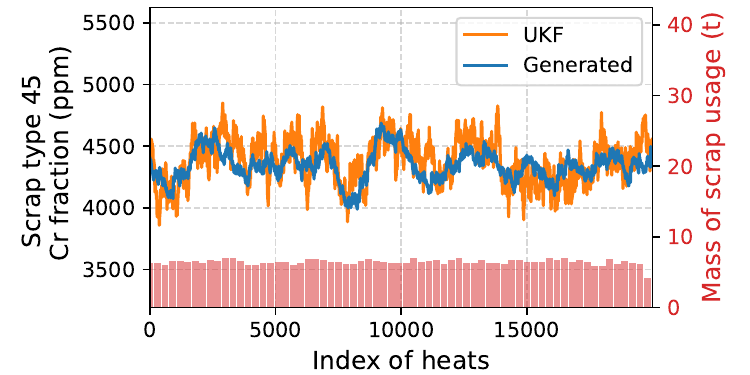}
    \end{subfigure} 
    \begin{subfigure}[b]{0.45\linewidth}
        \includegraphics[width=0.7\linewidth]{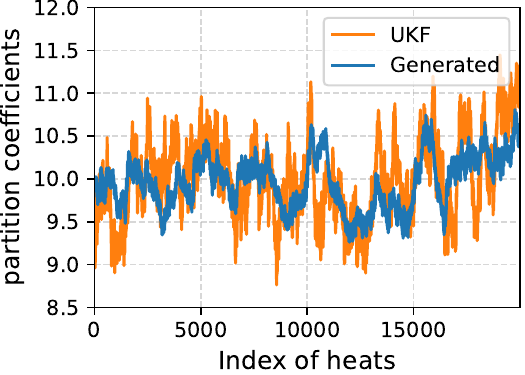}
    \end{subfigure} 
    \caption{Effects of $10\%$ measurement noise on iron oxide on the estimated partition coefficients and element fraction of scrap types.
    The left figure shows the estimated Cr fraction of scrap type $45$ and the right figure shows the estimated partition coefficient.
    }    
    \label{fig:KF_cr_fl_feon_20}
\end{figure}

This expectation is confirmed by the results shown in Figure~\ref{fig:KF_cr_fl_feon_20}, where the model is tested with \( 10\% \)  Gaussian noise in \( f_{{\rm FeOn, slag}, t} \). Despite this significant level of distortion, the estimated Cr fractions in scrap and the corresponding partition coefficients remain closely aligned with the true generated values. According to Table~\ref{tab:sens_f_feon}, both the bias and variance of the predictions remain low, and no significant deviation from the ground truth is observed.

These findings indicate that the current formulation for modeling the partition coefficient \( \ell_X \) is sufficiently robust for practical applications. This robustness is particularly beneficial, as direct measurement of the iron oxide fraction in slag is often subject to uncertainty, relying on laboratory analysis or indirect estimates. The model's ability to tolerate such noise without compromising estimation accuracy enhances its applicability in real-world steelmaking environments, where measurement imperfections are unavoidable.

Overall, the results suggest that while accurate data is always preferable, the estimation of partition behavior through a simple linear model using noisy \( f_{{\rm FeOn, slag}, t} \) remains effective and stable. This reduces the dependency on high-precision slag composition measurements, potentially lowering operational constraints for real process monitoring and control.

\begin{table}[htbp]
\centering
\caption{Effect of measurement noise of iron oxide in slag on predicted Cr fraction in steel (ppm). The first row is the same as the second row in Table~\ref{tab:sens_m_scrap_rank}, representing the theoretical best-case results for this case. Since the true value of $f_{{\rm steel, Cr}, t}$ ranges between $200$ to $400\,$ppm, all results reported in the table are considered acceptable.}
\label{tab:sens_f_feon}
\begin{tabular}{SSS}
    \toprule
    {${\rm noise}\,(\vec m_{\rm slag})$} & {${\rm E}[\varepsilon(f_{{\rm steel, Cr}, t})]$} & {$\sigma[\varepsilon(f_{{\rm steel, Cr}, t})]$} \\
    \midrule
    {$ 0\%  $}     & {$ -0.01 $}  &  {$ 5.39 $} \\ 
    {$ 5\%  $}     & {$ -0.30 $}  &  {$ 5.07 $} \\ 
    {$ 10\% $}     & {$ 0.03 $}  &  {$ 5.29 $} \\ 
    {$ 20\% $}     & {$ -0.32 $}  &  {$ 5.34 $} \\  
    \bottomrule
\end{tabular}
\end{table}

\section{Conclusion and Discussion}
\label{s:conclusion}
This study presented both linear and nonlinear state space models for estimating scrap composition in steelmaking, using mass balance equations relevant to Electric Arc Furnace (EAF) and Basic Oxygen Furnace (BOF) processes. A modified Kalman filter was applied for the linear model (with Cu as an example), and an Unscented Kalman Filter (UKF) was used for the nonlinear model (using Cr).

The primary focus was to evaluate the sensitivity of these models to measurement noise in key process variables, including steel mass,  steel composition, input mass of scrap types,  slag mass, and iron oxide fraction in slag.
The results indicate:
\begin{itemize}
    \item Errors in steel mass and composition can be effectively handled as observation noise.
     \item The model is relatively robust to noise in scrap input mass.
     \item The predicted elemental fraction in steel is quite robust to input mass matrices with different rank and condition number. However, accurate estimation of the elemental fraction of a specific scrap type is more likely when it is used more often and in greater amounts.
    \item The model is also sensitive to errors in slag mass, which can significantly affect the estimated fraction of scrap types.
    \item Errors in the iron oxide fraction have minimal impact, due to compensation by model parameters.
\end{itemize}

In summary, the model performs reliably in predicting the elemental fraction in steel when measurement errors are within $20\%$. As for the estimation of fraction of scrap types, the model is more sensitive. Among all inputs, accurate estimation of slag mass and accurate measurement of input scrap mass are most critical for model performance.

\section*{Acknowledgment}
The research leading to these results has been performed within the CAESAR project (\href{https://caesarproject.eu/}{https://caesarproject.eu/}) and received funding from the European Commission’s Horizon Europe Programme under grant agreement n° 101058520.

\bibliographystyle{plain}
\bibliography{reference}
\end{document}